\documentclass[preprintnumbers, floatfix, preprintnumbers, letterpaper, superscriptaddress,nofootinbib, onecolumn]{revtex4}
\pdfoutput=1
\usepackage{graphicx}
\usepackage{microtype}
\usepackage{amsmath}
\usepackage{amssymb}
\usepackage{subfigure}
\usepackage{hyperref}
\usepackage{url}
\usepackage{xcolor}
\usepackage{color}
\usepackage{mathrsfs}
\usepackage{calrsfs}
\usepackage{amsfonts}
\usepackage{latexsym}
\usepackage{ragged2e}
\usepackage{epsfig}
\usepackage{textcomp}
\usepackage{eufrak}
\usepackage{tabularx}
\usepackage{eucal}
\usepackage{epsfig}
\usepackage{amsmath} 
\usepackage[utf8]{inputenc}
\usepackage{fourier} 
\usepackage{array}
\usepackage{makecell}
\usepackage{longtable}

\definecolor{vividviolet}{rgb}{0.62, 0.0, 1.0}
\definecolor{amaranth}{rgb}{0.9, 0.17, 0.31}
\definecolor{palatinateblue}{rgb}{0.15, 0.23, 0.89}
\definecolor{brightpink}{rgb}{1.0, 0.0, 0.5}
\definecolor{cornflowerblue}{rgb}{0.39, 0.58, 0.93}
\definecolor{deepcarminepink}{rgb}{0.94, 0.19, 0.22}
\definecolor{radicalred}{rgb}{1.0, 0.21, 0.37}

\hypersetup{ linktoc=all,
	colorlinks, linkcolor={palatinateblue},
	citecolor={brightpink}, urlcolor={amaranth}
}

\graphicspath{{Images/}}

\newcommand\bfsigma{{\boldsymbol{\sigma}}}
\newcommand\tr{\mathop{\rm tr}\nolimits}

\renewcommand{\d}[1]{\ensuremath{\operatorname{d}\!{#1}}}

\def\sideremark#1{\ifvmode\leavevmode\fi\vadjust{\vbox to0pt{\vss
			\hbox to 0pt{\hskip\hsize\hskip1em
				\vbox{\hsize1.3cm\tiny\raggedright\pretolerance10000
					\noindent #1\hfill}\hss}\vbox to8pt{\vfil}\vss}}}%
\def\beq{\begin{equation}}
	\def\eeq{\end{equation}}

\begin{document}

\title{Observational backreaction in discrete black holes lattice cosmological models}

\author{Daniele \surname{Gregoris}}
\email{danielegregoris@libero.it}
\affiliation{Center for Gravitation and Cosmology, College of Physical Science and Technology, Yangzhou University, \\180 Siwangting Road, Yangzhou City, Jiangsu Province  225002, China}
\affiliation{School of Aeronautics and Astronautics, Shanghai Jiao Tong University, Shanghai 200240, China}

\author{Kjell \surname{Rosquist}}
\email{kr@fysik.su.se}
\affiliation{Department of Physics, Stockholm University,  106 91 Stockholm, Sweden}

\begin{abstract}
Applying the Sachs formalism, the optical properties encoded in the distance modulus are studied  along curves exhibiting local rotational symmetry  for some closed  inhomogeneous cosmological models whose mass content is discretized by Schwarzschild-like sources. These models may challenge the concordance model in its use of the distance modulus data of type Ia supernovae, because they do not violate any energy condition. This result relies only on the symmetry properties considered,  and not on the way in which the mass is discretized. The models with different number of sources are then compared among themselves and with a  Friedmann-Lemaitre-Robertson-Walker model with the same total mass content by introducing a compactness parameter. The analysis shows that {\it observational backreaction} occurs because increasing the number of sources the features of a universe with  a continuous matter distribution
are not recovered. Our models are shown to exhibit a non-trivial relationship between kinematical, dynamical and observational backreactions,  the kinematical one  being asymptotically decreasing while the latter two are present. Furthermore, the electric part of the Weyl tensor contributes to the luminosity distance by affecting the evolution of the scale factor, while the magnetic part has an indirect role by  affecting only the evolution of the former. 
\end{abstract}

\maketitle

\section{Introduction}
The presence of astronomical structures has been largely ignored in the modeling of the large-scale evolution of the Universe for two reasons: the cosmological principle and the non-linearity of the Einstein equations. The former was proposed with a physical motivation stating that the observer performing the measurements does not occupy a preferred position in space,  implying consequently that the real Universe is homogeneous and isotropic on large enough scales; the latter instead represents a mathematical technical challenge: analytical solutions of the gravitational field equations can be obtained only by assuming simplifications and/or imposing some symmetries to the problem from the start because they constitute a system of coupled non-linear partial differential equations.

When we try to account for the role of small-scale inhomogeneities, the core of the task is the explicit evaluation of the backreaction of such lack of homogeneity: this term quantifies the deviation of the large-scale evolution of an inhomogeneous Universe and a homogeneous one with the same total matter content. Cosmological backreaction is classified into kinematical, dynamical and observational backreaction, where the first concerns the initial data in the Cauchy formulation of General Relativity and requires only knowledge of the solution of the constraint part of the field equations;  the second refers to the different possible time evolutions of the systems, while the third is connected to the observational quantities that astrophysical missions can actually measure \cite{re2011}. The literature does not provide any general theorem relating the amount of one of them to the amount of the others,  and thus they must be evaluated explicitly for each system under examination. Moreover,   although one of the aforementioned kinds of backreaction may asymptotically decrease as the number of sources is increased, this does not guarantee a similar behavior for the others \cite{acc}. 

Among the many possible frameworks for describing the local inhomogeneous nature of the Universe we live in, one route consists in considering a regular lattice of black holes arranged on a 3-sphere \cite{cli1,cli2,cli3,cli4,be1,be2,be3,be4,irr,ben}.  The main numerical and analytical achievements obtained within this class of models in relativistic cosmology have been recently reviewed in \cite{re1}. In fact,  this choice exhibits many positive features: even if truly locally inhomogeneous, the system approaches homogeneity when coarse-grained on larger scales, as the physical Universe does; furthermore,  the regularity of the lattice comes with a number of the mathematical symmetries, like the Local Rotational Symmetry (LRS)  along specific networks of curves and the reflection symmetry of some surfaces, which permit to restrict the number of nonzero kinematical and curvature variables all along the time evolution of the system. Moreover, the time-reversal symmetry arising in a closed model has been used for obtaining exact initial data. Thus, this family of models falls inside the usual line of thinking adopted so far in theoretical cosmology which requires the reduction of the group of symmetry step by step. The kinematical and dynamical backreactions (the latter one along some specific curves in which the Einstein equations can be exactly integrated) have been already discussed in the previously cited references which have compared and contrasted the discrete models we will study in this paper to a homogeneous and isotropic Friedmann-Lemaitre-Robertson-Walker model with closed topology and the same total matter content pictured as a smooth dust fluid. The analysis of the initial data  suggests that kinematical backreaction asymptotically decreases when the number of the masses (Schwarzschild-like black holes) entering the configuration is increased, while, on the other hand, the time evolution of the discrete and fluid models is completely different. In particular, the lattice model suggests that a negative deceleration parameter may be possible without violating the energy conditions, i.e. without requiring the presence of any exotic fluid permeating the Universe, the dark and phantom energies being examples. This  particular case  explicitly confirms what we already stated: even having information about one kind of backreaction we can not infer the latter without explicitly computing it. In particular, two cosmological models governed by analogous dynamical equations do not  necessarily have analogous observational relations:  the detailed analysis of the observational backreaction for the black-hole lattice model is then in order.   

In this report we quantify the observational backreaction in a family of inhomogeneous cosmological models based on regular discrete black hole lattices providing another fundamental characterization of those models. This not only complements but also improves on previous studies for at least two other reasons. The first is that this analysis is not affected by the presence of the caustics (a non-physical  singularity in  the coordinate system) which played an important role in the quantification of the dynamical backreaction in \cite{cli2}. The second is that in the current concordance model of cosmology, the $\Lambda$CDM  model ($\Lambda$-Cold-Dark-Matter model), the trajectories of the motion of photons are affected by an {\it averaged spacetime curvature} which is the same at all spatial points. This is in contrast to the real Universe where instead the motion of the light is affected  by the {\it true curvature} which depends on the actual presence or absence of mass densities (generated by astronomical objects like galaxies, clusters of galaxies, gas clouds) along the line of sight \cite{peebles}. This latter case occurs in the inhomogeneous cosmological models like the one we will adopt in what follows. Intuitively, although accounting for  the  role of  astronomical structures does not completely solve the {\it  dark energy problem}, it is conceivable that it  will at least bring a correction to the amount of dark energy needed to account for the available astrophysical datasets, as already pointed out in \cite{corr}. This kind of numerical corrections to the values for the cosmological energy parameters can not be ignored nowadays because of the remarkable improvements in the cosmological measurements (see however \cite{Nielsen_etal:2016} for a critical evaluation of the supernova data).

The trajectories of photons encode information about the observational quantities of a given cosmological model through the luminosity distance equation. In a previous work \cite{clifton:2009} the propagation of the light beams was studied  in a black-hole lattice model assumed to represent an inhomogeneous Universe. They approximated the light rays at cosmological distances as geodesics in a Schwarzschild background nearby each mass and then matched the geodesics at the cell boundaries.
Now we will consider a different and improved approach using the lattice cosmology model: we will follow the light beam along edges 
(which are LRS curves) of cells in those configurations which admit contiguous edges between neighboring cells \cite{cli2}. Thanks to our formalism we will draw conclusions about the properties of the full spacetime without the need of any approximation because symmetric subspaces are totally geodesic and consequently a geodesic along an edge is also a geodesic of the full manifold \cite{rindler}.  In particular, we will show that the optical properties of those configurations depend on the local rotational symmetry of the curves along which the photons are moving, but not on the distribution of the massive sources inside the different discrete configurations. This work extends our previous one \cite{cli2} where we focused our attention only on the sign of the  deceleration parameter along the edge while here we consider the redshift as being more physically relevant in terms of observational relations.

The present paper is organized as follows. After a review in Sect.~\ref{sII} of the most important features of the inhomogeneous discrete cosmological model we will adopt, we will derive the Sachs equations governing the luminosity distance.
The integration of these equations allows a comparison  against the most recent type Ia supernova data collection in Sect. \ref{sIII}. In the same section we compare and contrast the luminosity distance in our models with other cosmological models discussed in the current literature. In Sect. \ref{sIV} we investigate the limiting case of a universe with infinitesimally close mass sources by exploiting the compactness parameter. Finally we comment our findings in Sect.~\ref{sV}.

\section{Black-hole lattices in the framework of cosmological relativistic modeling: a review} \label{sII}
In this section we review the most relevant mathematical properties of the {\it black hole lattice universes} that will be needed in the remaining part of our paper. For details the reader is referred to \cite{cli1,cli2,cli3,cli4}. Since we are constructing a regular tessellation of the 3-sphere in terms of Schwarzschild-like masses, only certain configurations admit an exact solution to the initial data equations. They are constituted by 5, 8, 16, 24, 120, and 600 sources respectively, and they can be interpreted as the solutions to the generalized Platonic problem of inscribing regular solids into a sphere, which can here be envisaged as living in an imagined  4-dimensional environment \cite{coxeter}. These initial data have been obtained at the moment of maximum expansion of the configuration of our closed model exploiting the time-reversal symmetry occurring there. Adopting spherical coordinates $(t,\,\chi,\, \theta,\, \phi)$ the solution reads
\beq
\label{metric0}
ds^2=\psi^4 (d\chi^2 + \sin^2 \chi d\theta^2 + \sin^2 \chi \sin^2 \theta d\phi^2 ) \,,
\eeq
where the conformal factor
\beq 
 \psi=\psi(\chi, \theta,\phi)=\sum_{i=1}^{N} \frac{\sqrt{m_i}}{  f_i(\chi, \theta, \phi)}
\eeq
depends on the values $m_i$ of the \emph{mass parameters}\footnote{Denoted ``effective mass" in \cite{cli1}.} of the $N$ sources, here taken to be all equal to each other, setting $m_i = m$ in what follows. The dependence of the conformal factor on the locations of the masses is specified by the functions $f_i(\chi, \theta, \phi)$. As an exemplification, in the 16-cell the sources are located at the centre of the cells whose positions in the coordinate system $w=\cos\chi$, $x=\sin\chi\cos\theta$, $y=\sin\chi \sin\theta\cos\phi$, $z=\sin\chi \sin\theta\sin\phi$ are given by all the permutations of $(\pm1,\pm1,\pm1,\pm1)$. In such a configuration the functions entering the conformal factor read as:
\beq
f_i=\sqrt  {  2\left(  1 \pm \frac12 \cos\chi \pm \frac12 \sin\chi \cos\theta \pm 
     \frac12 \sin\chi\sin\theta \cos\phi \pm
     \frac12  \sin\chi\sin\theta \sin\phi \right) } \,, \qquad i=1,...,16\,,
\eeq
where all the possible sign combinations must be considered.
Moreover, the mass parameter $m$ entering the conformal factor is related to the proper mass of the source $m_{\text{p}}$ which takes into account the energy of the gravitational field of all the other sources of the configuration according to the following conversion factors \cite{brill,cli1}
\begin{eqnarray}
&&\frac{m_{\text{p}}}{m}=0.20\,, \qquad \frac{m_{\text{p}}}{m}=0.11\,, \qquad \frac{m_{\text{p}}}{m}=0.045 \\
&&\frac{m_{\text{p}}}{m}=0.029\,, \qquad \frac{m_{\text{p}}}{m}=0.0052\,, \qquad \frac{m_{\text{p}}}{m}= 0.0010\nonumber
\end{eqnarray}
for the configurations with 5, 8, 16, 24, 120, and 600 masses respectively.
The dynamical equations governing the evolution of the LRS curves of our lattices (examples of which are the edges of the cells, the diagonals of the cells and the curves connecting two masses and passing through the centre of a cell face) can be fully written in terms of four quantities, the rate of expansion, one  component of the shear tensor, one component of the electric Weyl tensor and one spatial derivative component of the magnetic Weyl tensor. We stress the fact that no matter field  is present in our models,  and that the cosmological constant is set to zero as well. Our black hole configurations can be divided into two subclasses: the ones admitting contiguous edges (like the ones with 16, 24 and 600 masses) and the ones with non-contiguous edges (the remaining ones with 5, 8 and 120 masses) between two adjacent cells. The meaning of {\it contiguous edge configuration } is that the edge of a cell is superposed to the edge of a neighboring cell when prolonged in such a way that they constitute a smooth continuous curve at the cell corner which plays the role of a  matching point. Moreover, only in the contiguous-edge configurations the local rotational symmetry is a symmetry of the full lattice, while in the other previous mentioned cases it has only the range of a cell (see Table 3 in \cite{cli3}). Therefore,  since the light beams emitted by the supernovae  (on which the observational data considered in this paper are based) are required to have crossed a suitable fraction of the Universe to allow a cosmological interpretation of the model, we will restrict our attention to the configurations with contiguous edges.

We parametrize the edge curve with the $\chi$ coordinate and with the values of the coordinates $\theta$ and $\phi$ staying fixed.
Neglecting the role of the magnetic Weyl tensor, the solution for the metric in a synchronous reference frame can be given in a parametric form as \cite{cli2}:
\begin{eqnarray}
\label{metric}
ds^2 &=& -dt^2 + S^2(t,\chi) d\chi^2 \,, \qquad S(t, \chi)=  a_{||} \psi^2   \\
\label{factor}
a_{||} &=& \frac{ a_{||0}}{2}(3-\cosh^2\eta -3\eta \tanh\eta) \\
\label{time}
t- t_0 &=& \frac{1}{ \sqrt{ ^\circ E_0}} \left(\eta +\frac12 \sinh(2\eta)  \right)\,,
\end{eqnarray}
where ${a_{||0}}=a_{||}(\eta=0)$ is the value of the scale factor at the moment of maximum expansion, $ t_0$ is a constant of integration such that $t= t_0$ on the time-symmetric initial hypersurface, and $^\circ E$ is the value of the nonzero independent component of the Weyl electric field with a subscript 0 signifying its value at the moment of maximum expansion. 
 These equations have been derived under the assumption that the gravitomagnetic effects are negligibly small. We will quantify the uncertainty which affects our fitting procedure due to such effects in Sect. \ref{numsec}.
Finally we notice that a caustic appears in the synchronous frame  when $a_{||}=0$ (see Eq.~\eqref{factor}), that is at $\eta_{\rm cau}= 0.73$. We will show explicitly in Sect. \ref{sIV} that the caustic does not affect our results because we can account even for the most distant supernovae before reaching it.
 
\section{Optical properties of the discrete universes}\label{sIII}
\subsection{The Sachs equations in the discrete inhomogeneous model}
In this section we will introduce the Sachs equations governing the optical properties of a given cosmological model (see e.g.\ \cite{peebles,perlick,sac}) using the ON (orthonormal) frame formalism \cite{uggla}. We will focus on the case of light rays travelling along an edge of the cells. The edges constitute symmetry curves with the smallest curvature,  being the most distant  of such curves from the center of the cell which can be thought of as a center of gravity in these models.\footnote{Note however that the cell centers do not represent physical points in the spacetime of a lattice model just as there is no physical center of gravity in a spacetime with a single Schwarzschild source.}
We specify the first two vectors of the ON frame by assigning $e_0=u$ where 
$u$ is the observer four-velocity and letting $e_1=n$ be a unit vector along the edge (which is parametrized by the coordinate $\chi$ at fixed $\theta$ and $\phi$). We then define the ray 4-vector by $k=u+n$.
The remaining frame vectors, $e_2$ and $e_3$, represent a spatial 2-space often referred to as the screen space \cite{perlick}. 
The optical properties of a light beam can be specified by the expansion $\hat \theta$ and the shear which is a symmetric tracefree tensor $\sigma_{\!\!AB}$ ($A,B= 2,3$) in the screen space having the form \cite{peebles,perlick}
\beq
 \bfsigma = (\sigma_{\!\!AB}) =
 \begin{pmatrix}
  \sigma_1 & \sigma_2  \\
  \sigma_2 & -\sigma_1   \\
 \end{pmatrix} \,.
\eeq
The Sachs equations for the cross section of a light beam are given by  \cite{peebles,perlick,sac,ellisww}
\begin{align}\label{sachs1}
 \frac{d\hat \theta}{d\lambda}+\hat \theta^2+ 2 \tr{(\bfsigma^2)}
  \,&=\,-\tfrac12 R_{ab}k^a k^b \\[3pt]
  \label{sachs2}
 \frac{d \sigma_{\!\!AB}}{d\lambda}+2\hat \theta \sigma_{\!\!AB}&=C_{c\{\!AB\}d} k^c k^d\,,
\end{align}
where $\lambda$ is the affine parameter along the null geodesics and
$R_{ab}$ is the Ricci tensor and $C_{c\{\!AB\}d}$ is the Weyl tensor with two indices projected into a symmetric tracefree screen space index pair. For a general symmetric second rank tensor $S_{ab}$, such a projection is given by \cite{1+1+2}
\begin{equation}
    S_{\{\!ab \}} 
     = \bigl(f_{(a}{}^c f_{b)}{}^d - \tfrac12 f_{ab} f^{cd}\bigr) S_{cd}
\end{equation}
where $f_{ab} = g_{ab} + u_a u_b - n_a n_b$ is the screen space metric in a 1+1+2 framework \cite{1+1+2}.


%
When we restrict ourselves to the explicit case of the edge of a cell, the simplifications we should implement are:
\begin{itemize}
  \item $R_{ab}=0$ \,; the edge being in vacuum
  \item Along the edge, the Weyl tensor has only the single nonzero component $^\circ E = E_{ab} n^a n^b$ \cite{cli4}
\end{itemize}
where $E_{ab}$ is the gravito-electric part of the Weyl tensor. In general, $E_{ab}$ has the three linearly independent parts $^\circ E$, $^\dagger E^a$ and $^\ddagger{E}_{ab} = E_{\{\!ab\}}$ in the notation of \cite{cli4}.
The form of the righthand side of Eq.~\eqref{sachs2} for a given gravito-electric tensor is
\begin{equation}
    C_{c\{\!AB\}d} k^c k^d = 2 E_{\{\!AB \}}\,.
\end{equation}
Already at this point we can observe some differences between our model and the concordance model of cosmology in which the Ricci tensor is nonzero and depends on the energy density and pressure of the fluid permeating the Universe and in which the Weyl tensor plays no role, the Friedmann-Lemaitre-Robertson-Walker metric being conformally flat.
Thanks to the above simplifications, the Sachs equations (\ref{sachs2}) reduce to 
\begin{equation}
\frac{\d\sigma_{\!\!AB}}{\d\lambda}+2\hat \theta \sigma_{\!\!AB}= 0 \,.
\end{equation}
Imposing then the initial condition $(\sigma_{\!\!AB})_{\text{in}} = 0$ we obtain $\sigma_{\!\!AB} = 0$
for the evolution of the shear components, as expected from the LRS property of the edge. Thus, the only non-trivial Sachs equation we must integrate reduces to
\beq 
\label{sachsfin}
\frac{d\hat \theta}{d\lambda}+\hat \theta^2=0
\eeq
with the initial condition $\hat \theta_{\rm in} \to \infty$. The solution is then given by
\beq
\hat \theta=\frac{c}{\lambda}\,,
\eeq
where $c$ is an integration constant. 
Introducing the proper beam area $A$ through the relation
\beq
\frac{dA}{d\lambda}=2\hat \theta A \,,
\eeq
the only non trivial Sachs equation is reduced to 
\beq
\frac{d^2 \sqrt{A}}{d\lambda^2}=0 \,,
\eeq
which shows that the proper beam diameter $d_A$ is a linear function of the affine parameter along the null geodesics:
\beq
\label{areadistance}
d_A\,=\,k_1 \lambda +k_2\,,
\eeq
where $k_1$ and $k_2$ are constants of integration. To obtain a meaningful observational relation we need now to relate the affine parameter $\lambda$ (a measure of the distance,  naively speaking) to the redshift $z$ by integrating the null geodesic equations. To reach this goal we start by introducing the redshift  \cite{peebles,perlick}:
\beq
1+z\,=\,-\frac {(u_i k^i)_{\rm S}}{(u_i k^i)_{\rm O}}\,,
\eeq
where O denotes {\it observer} and S {\it source}, and where we can choose as initial condition $(u_i k^i)_{\rm O}=-1$ without loss of generality; thus
\beq
\label{redshift}
1+z=\frac{dt}{d\lambda}\,.
\eeq
 The system of  geodesic equations  in the metric (\ref{metric}) valid along an edge is given by
\begin{eqnarray}
&&\frac{d^2t}{d\lambda^2}+S \dot S \left(\frac{d\chi}{d\lambda}\right)^2\,=\,0 \\
&&\frac{d^2\chi}{d\lambda^2}+2 \frac{\dot S}{S} \frac{dt}{d\lambda} \frac{d\chi}{d\lambda}+\frac{S'}{S}\left(\frac{d\chi}{d\lambda}\right)^2\,=\,0
\end{eqnarray}
where a dot denotes a derivative with respect to the coordinate time $t$,  and a prime with respect to the angular coordinate $\chi$. For the case we are interested in we must also implement the null condition $ds^2=0$. With a few algebraic manipulations it is possible to show that the two independent equations can be cast as
\begin{eqnarray}
\label{geodesict}
&&\frac{d^2t}{d\lambda^2}+\frac{\dot S}{S} \left(\frac{dt}{d\lambda}\right)^2\,=\,0 \\
&&\frac{dt}{d\lambda}\,=\, S \frac{d\chi}{d\lambda}\,,
\end{eqnarray}
where the explicit geodesic equation for $\chi$ is redundant.

For comparing the Sachs equations to the astrophysical data we must relate the affine parameter $\lambda$ to the redshift $z$. 
For the numerical treatment and the analysis of the observational relations, the geodesic equations should be handled as follows. We move to a first order system of ordinary differential equations  implementing the definition of the redshift (\ref{redshift}):
\begin{eqnarray}
\label{system}
&& \frac{dt}{d\lambda}\,=\, 1+z \nonumber\\
&& \frac{dz}{d\lambda}\,=\,-\theta_{11}(t,\, \chi) (1+z)^2 \\
&& \frac{d\chi}{d\lambda}\,=\,P(t,\, \chi) (1+z) \nonumber\,,
\end{eqnarray}
where $\theta_{11}(t,\, \chi)=({\rm ln}a_{||})_{,t}=H_{||}$ has been introduced in \cite[Eq.~(4.12)]{cli2}, while $P(t,\, \chi)=\frac{1}{S(t, \, \chi)}$ drives the redshift drift as seen by dividing side by side the second and the third equation.  We can note that only the parallel component of the Hubble function to the edge enters the equation,  as expected since only the projection along the line of sight should contribute to the redshift.
By dividing side by side the second and first equation of the previous system,  using the definition for $H_{||}$ and (\ref{factor}) we get
\beq
\label{redshiftbis}
a_{||}=\frac{a_{||0}}{2}(3-\cosh^2\eta -3\eta \tanh\eta)=\frac{1}{1+z}\,.
\eeq
 This result is in agreement with the general literature studying how  spatial anisotropies are related to the redshift. In fact, the redshift is a measure of the stretching that affects the photons' wavelength ${\bar \lambda}$ when crossing a distance $l$ caused by the expansion of the universe. The magnitude of this effect depends on the kinematical  variables via \cite{contel} (see also \cite[Eq.~(141)]{ellisvan}):
\beq
\frac{d{\bar \lambda}}{{\bar \lambda}}=\left(\frac{\Theta}{3} +\sigma_{ab}e^a e^b \right) dl=\left(\frac{\Theta}{3} +\sigma_{11} \right) dl=H_{||}dl\,.
\eeq
In the second step of the previous equation we have used that $e^a=e^1$ for accounting for the direction of propagation of the light beam, while in the last step we have implemented \cite[Eq.~(4.12)]{cli2}. 


We also find an analytical  parametric relation between the affine parameter $\lambda$ and the redshift $z$ in terms of the conformal time $\eta$ by means of (\ref{factor}) and (\ref{time}),  and considering $t=t(\lambda(\eta))$ in the first eq. of (\ref{system}). We get
\beq
\label{lambda}
\lambda=\frac{a_{||0}}{\sqrt{^\circ E_0}} \left[\frac98 \eta -\frac34 \eta \cosh(2\eta)+\frac78 \sinh(2\eta)-\frac{1}{32} \sinh(4\eta)   \right]\,.
\eeq
Applying Etherington's theorem \cite{Etherington} we move from the area distance (\ref{areadistance}) to the luminosity distance:
\beq
\label{DL}
d_L\,=\, (1+z)^2 d_A\,,
\eeq
and finally to the distance modulus,  which is defined as \cite{peebles}
\beq
\label{modulus}
\mu_0\,=\, 5 \,{\rm log}_{10} \frac{d_L}{d^{\rm Milne}_{L}}\,,
\eeq
where the luminosity distance for the Milne Universe is given by
\beq
d_L\,=\, z\left( 1+\frac{z}{2} \right)\,.
\eeq
Normalizing the affine parameter (since it is not a measurable quantity we can rescale and shift it) by fixing the constants $\lambda_1$ and $\lambda_2$ in such a way that $\lambda_{z=0}=0$ and $(\frac{d\lambda}{dz})_{z=0}=1$, which in particular eliminates the dependence on the electric Weyl tensor.  Inverting (\ref{redshift}) for $\eta$ as a function of $z$, and plugging it into (\ref{lambda}), and then into the distance modulus (\ref{DL})-(\ref{modulus}), we get an expression that depends on the physical quantity $z$ and on the only left free parameter  $a_{||0}$. This latter quantity will be estimated by comparing our model with the observational data. We stress that we will get the same result for the fitting procedure for all the black holes configurations under investigation because it depends only on the LRS symmetry of the edges and not on the mass distribution. Therefore, the electric Weyl field contributes to the luminosity distance through $a_{||}$, and the magnetic Weyl field indirectly by affecting the evolution of the former.

\subsection{Numerical analysis}
\label{numsec}
\begin{figure}
\begin{center}
\includegraphics[scale=0.7]{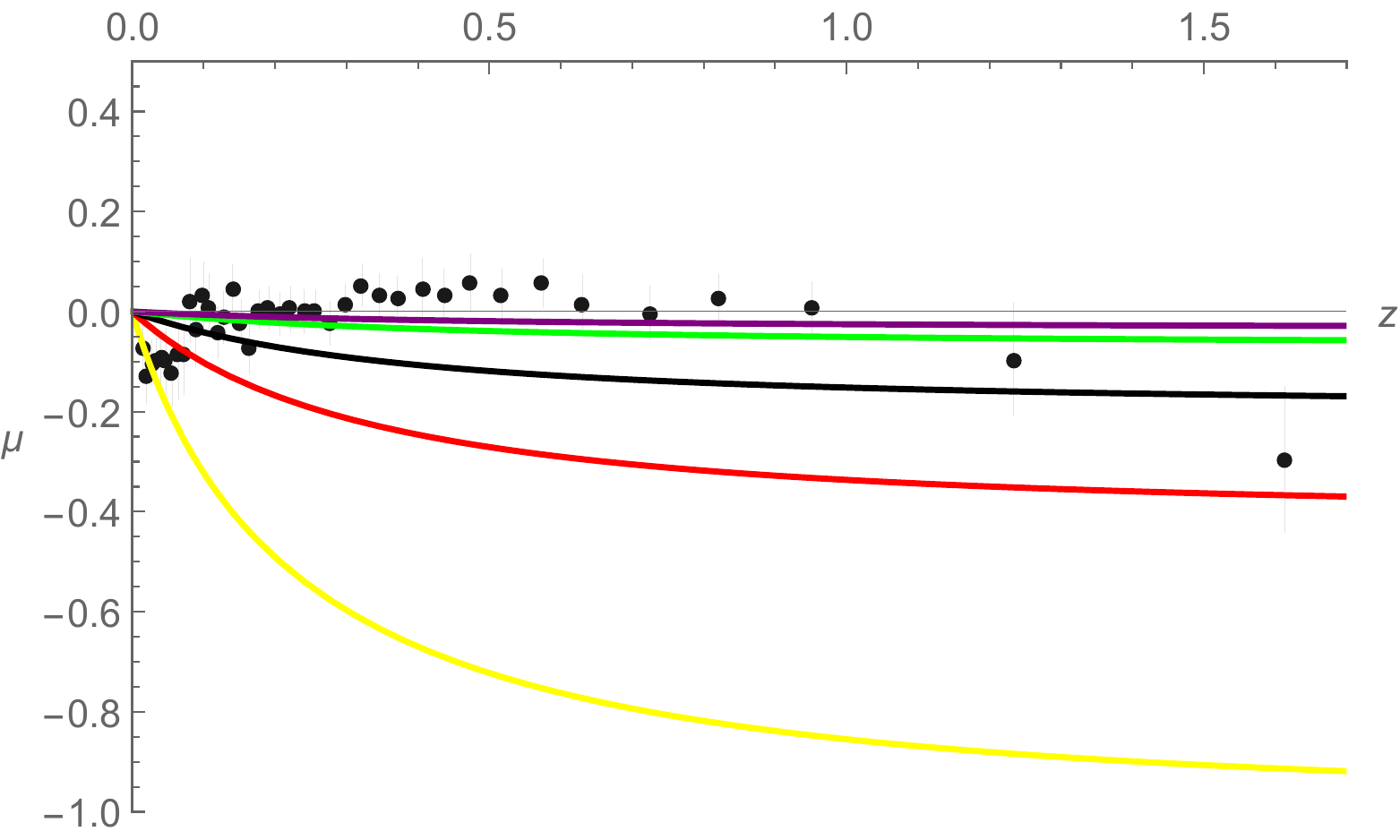}
\end{center}
\caption{ This figure is a comparison between our model, for some values of $a_{||0}$, and the data points of the distance modulus $\mu$ vs.~ redshift taken from \cite{super1,super2}. The $\mu$ values are computed from eqs (\ref{modulus}) and (\ref{eqmu}) given here for some values,  from bottom to top, $a_{||0}=1.1$  (yellow), $a_{||0}=1.3$  (red), $a_{||0}=1.6$  (black, best fitting case), $a_{||0}=2.3$  (green), and $a_{||0}=3.0$  (purple); the latter two cases are almost superposed to each other. For the sake of clarity in this figure we have decided just to exhibit the binned data points \cite{binned} obtained applying the BBC  method \cite{BBC}. }
\label{figmu}
\end{figure}

From now on a numerical treatment of our equations is required. For estimating the best fit value of the scale factor at the moment of maximum expansion we must test the theoretical expected value of the distance modulus 
\beq
\label{eqmu}
\mu \,= \, \mu_0 + m_b
\eeq
against the latest {\it pantheon} datasets obtained from the study of the type Ia supernovae \cite{super1,super2}. Numerically $m_b=25+X$, where the value of X is the absolute peak luminosity of a type Ia supernova which appears to be about $-1$ for the dataset considered here.  Since all the cosmologically meaningful curves $\mu=\mu(z)$ must fulfill $\mu(0)=0$,  we can fix the value of $m_b$ by requiring that the average value for $\mu(z)$ in the low-redshift regime should be small. A quantitative and model-independent procedure due by Goliath et al.  \cite{goliath} removes the dependence of the physical quantities characterizing the supernovae on the absolute magnitude of the supernovae sources since it is a factor of uncertainty. The procedure reads:
\begin{equation}
m_b = \frac{\displaystyle \sum_{i=1}^{n} \frac{\Delta}{\sigma_i^2}}
            {\displaystyle\sum_{i=1}^{n}\frac{1}{\sigma_i^2}} \,,\qquad
\Delta =  \mu_{\rm measured}-\mu_{0 \rm expected} \,,
\end{equation}
where $\sigma_i$ is the uncertainty in the measured quantities. For the best fit value (see next paragraph) of $a_{||0}=1.6$ we obtain $m_b=24$ in agreement with our previous statement. For estimating the best fit value for the free parameter $a_{||0}$ in our model we minimize the quantity
\begin{equation}
   \chi^2 = \sum_{i=1}^{n} \frac{(  \mu_{\rm measured}-\mu_{ \rm 
            expected})^2}{\sigma_i^2} \,. 
\end{equation}
The results of this procedure are displayed in Table \ref{table} where ``d.o.f.'' stands for the number of degrees of freedom, i.e.\ the number of available data points, which is 1048 for the {\it pantheon} dataset. Since the minimum value of $\chi^2$ is O(0) with the number of the degrees of freedom given by the number of measured data points, our model turns out to be a realistic picture of the real Universe as far as the type Ia supernova data are considered.
See also Fig.~\ref{figmu} where a graphical summary of the comparison between our model and the data points is given; in the figure, $\mu$ has been computed from eqs.~(\ref{modulus}) and (\ref{eqmu}) with  the choices, for graphical convenience,  of $a_{||0}=1.1$, $a_{||0}=1.3$, $a_{||0}=1.6$  (best fitting case), $a_{||0}=2.3$, and $a_{||0}=3.0$; the latter two cases result are almost superposed on each other.  For the sake of clarity in this figure we have decided just to exhibit the binned data points taken from \cite{binned} which have been obtained using the so-called BEAMS with Bias Corrections Core-Collapse (BBC in short) method \cite{BBC}.
 We also note that the {\it present time} value of the scale factor is the same for all configurations as we refer to it as a function of $\eta$ from (\ref{redshift}) which will correspond to $z=0$. In particular we can compute $\eta_{0}= 0.44  < \eta_{\rm cau} = 0.73$ (see Sect.~\ref{sII}) which confirms that our present analysis, contrary to the one in \cite{cli2}, is not affected by the occurrence of a caustic in the evolution of the spacetime.  However the estimated  physical time would be different for each configuration and different for each point along the edge due to the effects of the electric Weyl tensor in (\ref{time}) and it can be determined only up to the constant $t_0$ due to the gauge invariance of general relativity.

\begin{table}
\begin{center}
\begin{tabular}{ |l|l| }
\hline
  $a_{|| 0}$ & $\chi^2 /\text{d.o.f.}$ \\
  \hline
  1.2      &   3.85013      \\
  1.3     &     2.08722                  \\
    1.4  &      1.58911                    \\
1.5      &     1.44208                       \\
 1.6     &      1.41232                       \\
  1.7   &        1.42453                          \\
   1.8   &      1.45116                       \\
  1.9   &        1.48138                        \\
  \hline
\end{tabular}
\end{center}
\caption{ This table is a summary of the results of the optimization procedure for the $\chi^2$ with respect to the value of the scale factor at the moment of maximum expansion  $a_{||0}$. The abbreviation ``d.o.f'' represents the number of degrees of freedom, i.e. the number 1048 of the available data points.  }
\label{table}
\end{table}

 In order to estimate the magnitude of the uncertainty which affects the outcome of our fitting procedure  from neglecting possible gravitomagnetic effects (their role is indeed important in cosmology because they encode information about the gravitational interactions between astrophysical structures in motion with respect to each other) we apply the following procedure:
\begin{itemize}
\item We estimate the time interval we need to cover for taking into account all our datapoints by implementing the information that $\eta_{0}=0.44$ and the initial value of the gravitoelectric field, that we can obtain by looking at \cite[Fig.~(3)]{cli4},   into eq.~(\ref{time}). For getting a quantitative result we set $ t_0=0$, and we assume that the light beam is originated from the middle point of the edge.
\item We quantify the uncertainty on the estimate of the scale factor by plugging the previous information and the magnitude of the gravitomagnetic field, which can be obtained by looking at \cite[Fig.~(5)-Fig.~(7)]{cli4},  into \cite[Eq.~(74)]{cli4}.
\item We summarize the results of this procedure for the 16 and 24 cell models (these configurations come with contiguous edges which is an assumption required by the analysis performed in this manuscript) in Table \ref{error}.
\end{itemize}
The outcome of this analysis suggests that our fitting procedure can be trusted mainly for lattice models with a small number  of sources.

\begin{table}
\begin{center}
\begin{tabular}{ |l|l|l|l|l|l| }
\hline
  Lattice type & \, $^\circ E_0$  & \,\,\,\,\,\,\, $t$ & $^\circ {\rm curl \, H}_0'''$ & $^\circ {\rm curl \, H}_0'''''$ & $\frac{\Delta a_{||}}{ a_{||}}  $\\
  \hline
  16 masses    &  0.0170 & \, 7.2020 & \,\, -0.00017   & \,\,\,\,\, 0.0012 & 0.03    \\
  24 masses    &   0.0075  & 10.8430 &  \,\,\, 0.00015  & \,\,\,\, -0.0001 &  0.34            \\
    
  \hline
\end{tabular}
\end{center}
\caption{ This table reports on the level of uncertainty affecting our fitting procedure about the scale factor which has been estimated by implementing the magnitude of the gravitomagnetic effects into \cite[Eq.~(74)]{cli4}. In this analysis we are assuming that the light beam is originating from the middle point of the edge. We can appreciate the growth of the uncertainty level with the number of sources populating the lattice model in agreement with the discussion of  \cite{cli4}. }
\label{error}
\end{table}

Our model can be further  compared with other cosmological models currently adopted by the community comparing and contrasting the luminosity distances they are characterized by. The luminosity distance for our model has been derived in the previous paragraph, while for the other relevant cases the literature provides (consider eq. \cite[Eq.~(15.3.24)]{wein1972}) :

\begin{itemize}
\item Einstein-De Sitter model \cite{ein} (see also \cite[Eq.~(17)]{hogg}):
\beq 
\label{eds}
H_0 d_L(z) = 2  ( 1+z - \sqrt{1+z} )
\eeq
 \item Friedmann-Lemaitre-Robertson-Walker model \cite{peebles,perlick} (see also \cite[p.79]{goobar}):
\beq
\label{frw}
H_0 d_L(z)= \frac{1+z}{\sqrt{|\Omega_{k0}|}}  \sin\left[\int_0^z \frac{\sqrt{|\Omega_{k0}|} \, ds}{   \sqrt {  \Omega_{m0} (1+s)^3+ \Omega_{k0} (1+s)^2 +\Omega_\Lambda}}\right]\,,
\eeq
which corresponds to the standard model of cosmology $\Lambda$CDM if we pick $\Omega_\Lambda =0.69$, $\Omega_{k0}=-0.01$ (which implies the choice of the sine function  which corresponds to a closed universe),  and $\Omega_{m0}=0.30$. We note that this choice for the numerical values of the parameters delivers $\chi^2 / \text{d.o.f.} \sim 3$, which is higher than our best estimate for the discrete universe (however this set of cosmological parameters have been estimated also in light of the cosmic microwave background radiation, and of the baryon acoustic oscillations, and not only of the supernovae data).
\item Empty-Beam Approximation (EBA) for an Einstein-de Sitter universe  due to Zel'dovich \cite{ebaref} and Dyer-Roeder \cite{ebaref2, ebaref3}:
\beq
\label{eba}
H_0 d_L(z)= \frac{ 2(1+z)^2 } {5} \left( 1-\frac{1} {(1+z)^{5/2}} \right)\,.
\eeq
This EBA constitutes the closest approximation both to the luminosity distance that is obtained in a black hole lattice when numerical methods are employed \cite{be4},  and to the tidal effects on a light beam propagation due to gravitational lensing  when a stochastic Langevin approach is followed \cite{arenaobs}. Furthermore, it is consistent with supernovae data as well \cite{ebasn}. Eq.~(\ref{eba}) is just a particular case of \cite{ebaref2}:
\beq
\label{ebag}
H_0 d_L(z)=(1+z)^2 \int_0^z \frac{ds}{(1+s)^3 \sqrt{1+2q_0 s}}
\eeq
when the deceleration parameter is set to be $q_0=\frac12$. We recall that for a dust Friedman-Lemaitre-Robertson-Walker universe the deceleration parameter is related to the curvature and matter parameters $\Omega_k$ and $\Omega_m$ via $q=\Omega_k+\frac32 \Omega_m -1$  \cite{ellisvan}. Thus, explicitly this condition is equivalent to $\Omega_{k0}=\tfrac32(1-\Omega_{m0})$.
\item Milne model \cite{milneref}:
\beq
\label{milne}
H_0 d_L(z) = z  \left( 1+\frac{z}{2} \right) \,.
\eeq
\end{itemize}

\begin{figure}
\begin{center}
\includegraphics[scale=0.7]{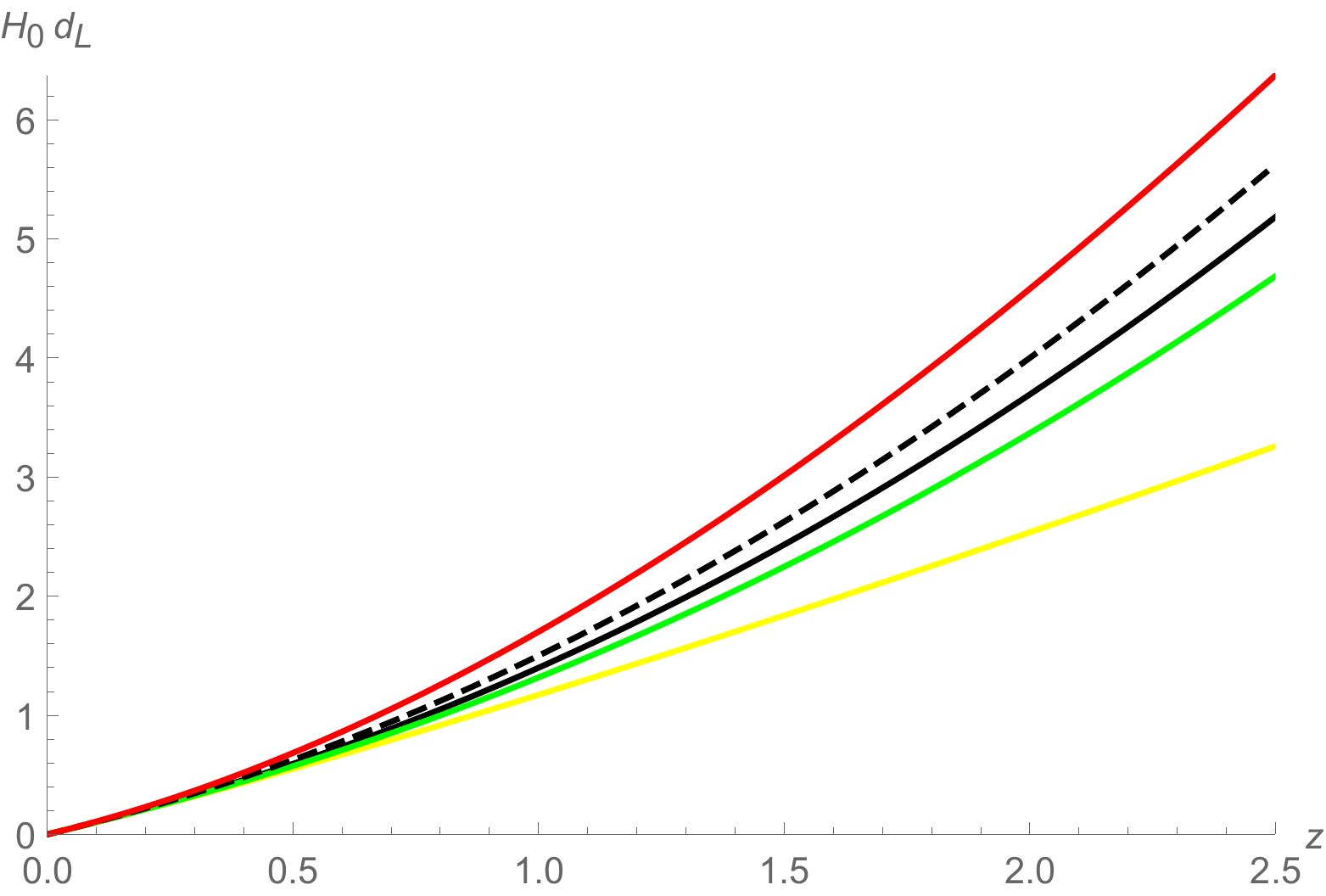}
\end{center}
\caption{ The figure is a comparison between the luminosity distances in the models, from bottom to top,  the Einstein-De Sitter model (yellow, eq.~(\ref{eds})), the particular case of the EBA given in eq.~(\ref{eba})  (green), the $\Lambda$CDM model (red, eq. (\ref{frw})), our model with the best fit choice $a_{||0}=1.6$ (black) and the Milne model (dashed, eq.~(\ref{milne})). The distance unit is the inverse cosmological constant $H_0^{-1}$.  The aforementioned EBA model is also considered the best approximation for describing the light propagation in a black hole lattice when numerical relativity methods are employed \cite{be4}.  }
\label{figdl}
\end{figure}

\begin{figure}
\begin{center}
\includegraphics[scale=0.7]{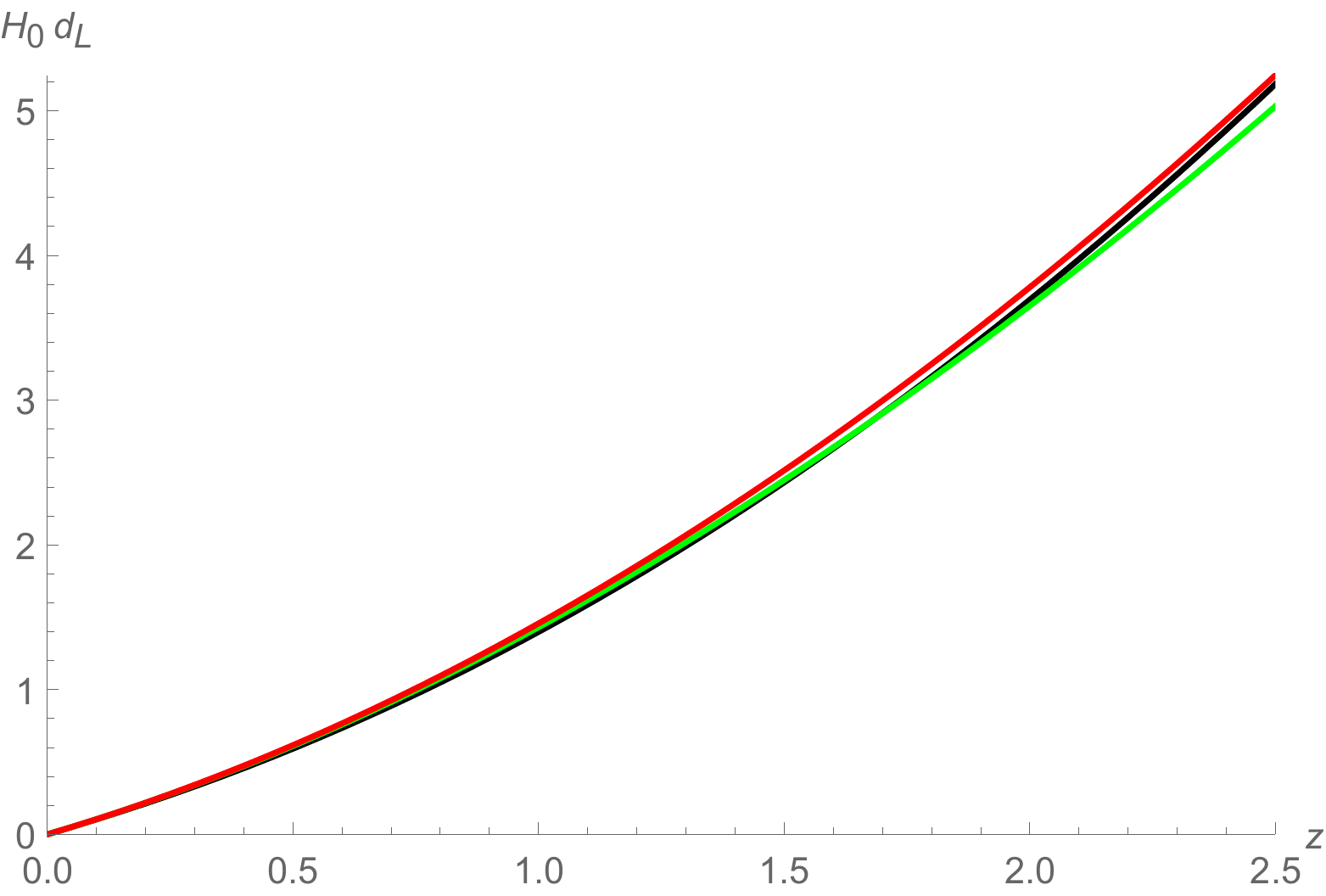}
\end{center}
\caption{The luminosity distance corresponding to our best fit black hole lattice model (black) is shown with the neighboring distances for EBA cases from eq.~(\ref{ebag}) with the choices $q_0=0.05$ (green) and $q_0=0.03$ (red). The close similarity is noteworthy.}

 
\label{fig4}
\end{figure}

\begin{figure}
\begin{center}
\includegraphics[scale=0.7]{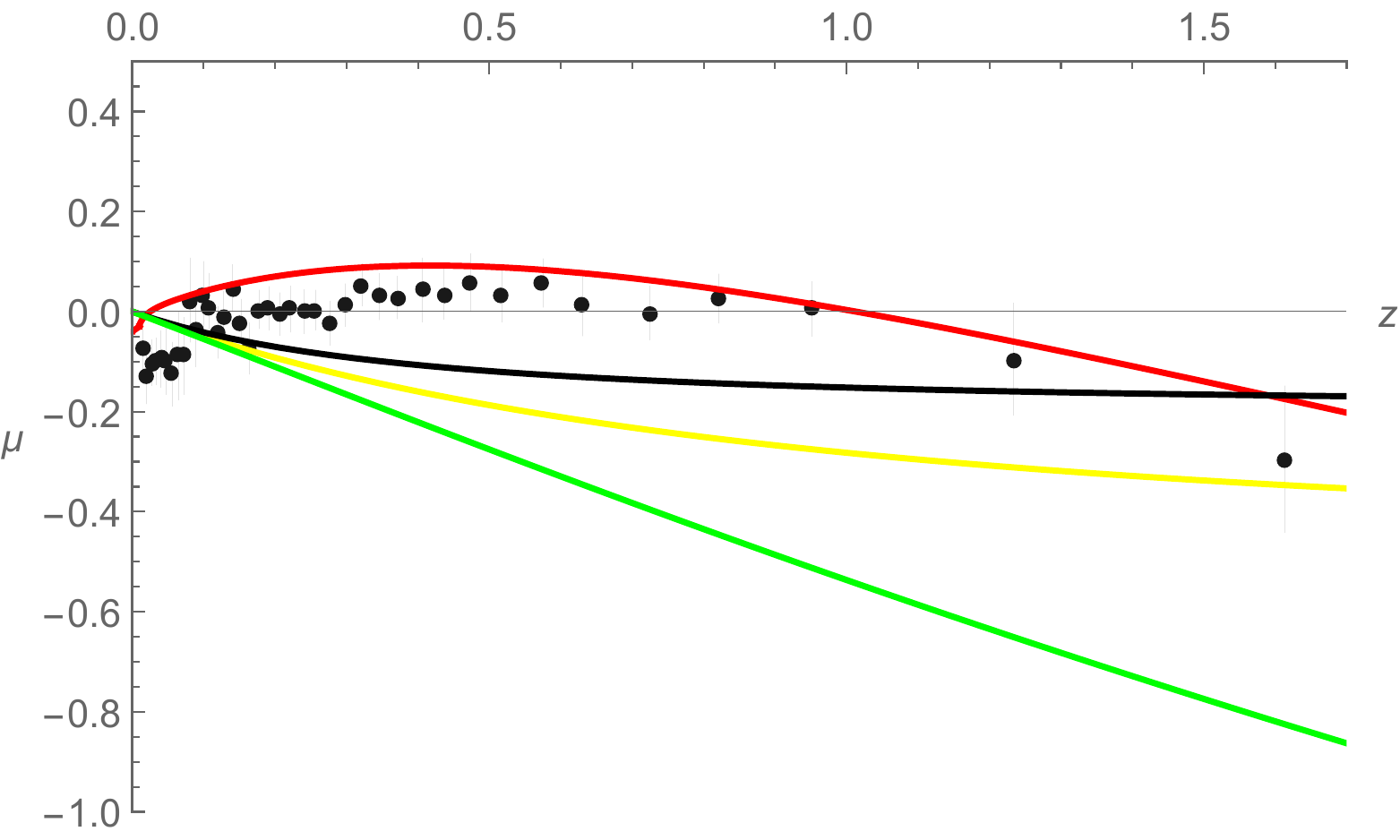}
\end{center}
\caption{ In this figure we compare our best fit lattice  model (black curve) with the binned data. Curves for some other models are also shown for comparison. From bottom to top, the green, yellow and red curves correspond respectively to the cases of  Einstein-de Sitter, EBA as given in eq.~(\ref{eba}), and  $\Lambda$CDM.}
\label{compmu}
\end{figure}

In Fig. \ref{figdl}, the luminosity distance,
obtained in our model for the best fit choice $a_{||0}=1.6$, is compared with the Einstein-de Sitter model, the $\Lambda$CDM model, the specific EBA model given in eq.~(\ref{eba}), and the Milne model. For a clearer comparison with the literature we display in Fig. \ref{fig4} only the curves corresponding to  our black hole lattice model, and two EBA cases corresponding to eq.~(\ref{ebag})  with $q_0=0.05$ and  $q_0=0.03$, respectively,  which result almost superposed. This implies that the closest approximation to our analytical treatment is provided by the EBAs, i.e. our analysis is in agreement with the numerical treatment presented in \cite{be4}  and with the stochastic description of a tidal gravitational lens effect on the trajectory of a light beam \cite{arenaobs}. In Fig. \ref{compmu} we display a comparison between the above mentioned models and the binned data.

One interpretation of the results shown in  Figs.~\ref{figmu} and \ref{figdl} is that our vacuum discrete cosmological model to some extent mimics Friedmann-Lemaitre-Robertson-Walker models with dark energy.
This conclusion depends on the optical properties of the models when tested against the supernova data and we did already obtain a similar result in \cite{cli2} when we computed the deceleration parameter. A key point of these results is that they hold for a vacuum spacetime without violating any energy condition and it can be interpreted as a direct consequence of using models which are both inhomogeneous and anisotropic while still possessing an approximate global isotropy. This calls for a more careful interpretation of the Hubble diagram because it suggests that dark energy may  in truth be only an {\it interpretative aspect} of the current model of cosmology,  and its indirect evidence from the supernova data may not constitute a solid physical result as some authors have been pointing out recently \cite{hu1,hu3, Nielsen_etal:2016}. In fact, even before those observations it was argued that inhomogeneities along the line of sight would affect the motion of the light rays coming from an astrophysical source \cite{hu2}.  Post-Newtonian simulations applied to the cosmological modeling have shown as well that the amount of dark energy inferred from the Hubble diagram may disagree with the one suggested by the Friedmann-Lemaitre-Robertson-Walker model due to approximations of the light beams trajectories \cite{viraj}.
As a further result, the figures also show that observational backreaction is present in our model since its optical properties are clearly not those of a dust dominated Friedmann-Lemaitre-Robertson-Walker Universe, and we recall that dynamical backreaction is as well an important feature of black hole lattices \cite{cli2}, while kinematical backreaction is negligible \cite{cli1}.

\section{The limiting case of infinitesimally close sources} \label{sIV}
The numerical analysis and the fitting procedure discussed in the previous section do not rely on the particular mass distribution considered, but only on the symmetry properties of the submanifold on which the light rays are assumed to travel. For a better understanding of how the inhomogeneities affect the optical properties of our model we  introduce the dimensionless {\it compactness parameter} 
\beq
\label{compact}
C = \frac {m_{\text{p}}} {l} \propto \rho l^2 \,,
\eeq
where $m_{\text{p}}$ is the proper mass of each source, $l$ is the length of a reference curve, being the edge in our case and $\rho$ is the mass density. We remark that in the second step of the previous equation we have only proportionality and not equality due to different geometrical factors relating the mass to the density for the different configurations considered in this paper. In general such a length is a function of the cosmic time $t$ as displayed in  \cite[Figure~10]{cli2},  but here we consider it at its present value, i.e. at redshift $z=0$. Thus, for an edge aligned with the $\chi$ direction we can obtain
\beq
l = \int_{\chi_1}^{\chi_2} \sqrt{ g_{\chi\chi}} d\chi = a_{|| {0}}  \int_{\chi_1}^{\chi_2} \psi^2(\chi)  d\chi \,,
\eeq
where $\chi_1$ and $\chi_2$ are the coordinates of the cell corners which constitute the edge endpoints, and $a_{|| {0}}$ is the present value of the scale factor which is the same for all our configurations. 
Numerically we get:
\beq
C_{16}=1.74 \cdot 10^{-4}\,, \qquad C_{24}=7.13 \cdot 10^{-5}\,, \qquad  C_{600}=3.56 \cdot 10^{-8}\,.
\eeq
We can note that once the reference length is assumed to be evaluated at a specific moment, it does not matter at which time we estimate the ratio between the compactness parameters. Observing that the compactness parameter is proportional to the square of the scale factor, and from the relation above which confirms that the coarse-grained Friedmann-Lemaitre-Robertson-Walker limit corresponds to taking $l \to 0$ (since the value of the compactness parameter is decreasing when the number of mass sources is increasing) we understand that the continuum limit \cite{cont} corresponds to taking $a_{|| 0} \to 0 $ in the  luminosity distance. 

\begin{figure}
\begin{center}
\includegraphics[scale=0.7]{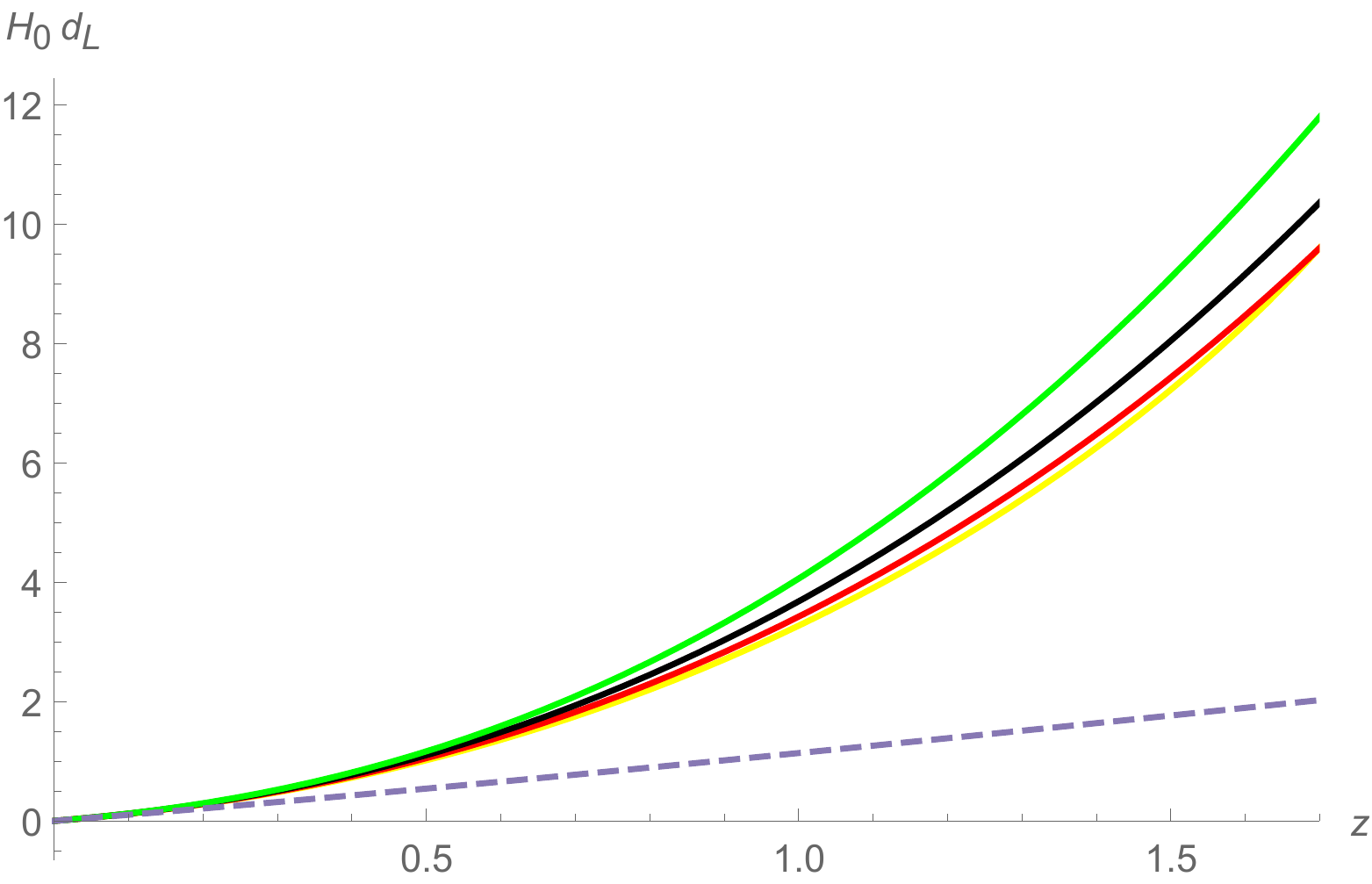}
\end{center}
\caption{The luminosity distance corresponding to the black hole lattice model is shown assuming  $a_{|| 0} =0.35 $ (yellow curve), $a_{|| 0} =0.30 $ (red curve), $a_{|| 0} =0.25 $ (black curve), and $a_{|| 0} =0.20 $ (green curve). Thus, the discrete universes exhibit observational backreaction because decreasing the value of $a_{|| 0}$ (which corresponds to taking the continuum limit for their matter content) their optical properties depart from those of a closed dust Friedmann-Lemaitre-Robertson-Walker universe (dashed curve).}

 
\label{fig5}
\end{figure}

Thus, for establishing whether observational backreaction arises, we must investigate if the luminosity distance (\ref{DL}) for the lattice models approaches the one of a closed dust Friedmann-Lemaitre-Robertson-Walker universe in the limit $a_{|| 0} \to 0 $. The latter is given by the formula
\beq
\label{frwc}
H_0 d_L(z)= \frac{1+z}{\sqrt{|-1/6|}}  \sin\left[\int_0^z \frac{\sqrt{|-1/6|} \, ds}{   \sqrt {  7/6 (1+s)^3-1/6 (1+s)^2 }}\right]\,,
\eeq
because in units of $H_0=1$ the curvature parameter of a closed Friedmann-Lemaitre-Robertson-Walker universe reads $\Omega_{k0}=-\frac{k}{6 H^2_0}=-\frac{1}{6}$, and the corresponding matter abundance parameter follows from the Gauss constraint $\Omega_{m0}+\Omega_{k0}=1$. According to our numerical analysis reported in Fig. \ref{fig5} observational backreaction {\it does} arise in our model because the optical properties of the discrete universes do not approach the one of a closed dust Friedmann-Lemaitre-Robertson-Walker cosmology as $a_{|| 0} \to 0 $, and actually they depart from it. We stress that observational backreaction is defined with respect to the maximally-symmetric counterpart, and should not depend on specific data used as a proxy, and therefore it is irrelevant whether the closed dust Friedmann-Lemaitre-Robertson-Walker universe can account for the supernovae datasets or not.

\section{Conclusion}\label{sV}
The most widely adopted technique for addressing the formation of cosmological structures is perturbation theory applied to a homogeneous and isotropic Friedmann-Lemaitre-Robertson-Walker spacetimne \cite{infla}. However, this framework comes with a variety of problems ranging from the absence of a mechanism for the formation of primordial black holes, up to the absence of a clear physical mechanism for exiting the early inflationary epochs of the universe \cite{dolgov}. In the light that the Copernican principle at the core of the current cosmological modelling is nothing else than a philosophical requirement \cite{ellis}, it has been argued that inhomogenous solutions of the Einstein equations may address these open problems \cite{bolejko}. On the other hand, the cosmological principle is also challenged on large scales by the observations dealing with the number counts which exhibit anisotropies on larger length scales than those of the cosmic microwave background \cite{ani1}. Furthermore,  also the study of  the energy jets emitted by quasars suggests the presence of anisotropies on cosmological scales which cannot be accounted for within the standard model of cosmology \cite{ani2, ani3, ani4}. Thus, a modification of the current cosmological paradigm seems to be  in order.
Of course the newly proposed general relativistic metrics must fulfill also all the other observational requirements, the luminosity distance of the supernova being one example. In this paper we have checked that black hole lattices models can constitute a realistic picture of the universe with respect to this latter requirement exploiting the discrete symmetries they admit. In particular,  we have identified some totally geodesics submanifolds which has allowed  us to trace the light rays analytically. Interestingly, one improvement is that these models do not require the existence of mysterious and undetected fluids as the $\Lambda$CDM model does,  as far as the supernova data interpretation is considered.  These models differ from the current cosmological framework for a number of reasons, as they are in vacuum with a vanishing Ricci tensor,  but they are affected by  a nonzero spacetime shear and Weyl tensors, which make the reconstruction of their optical properties a non-trivial problem that we have addressed in this paper. This work opens then the path for a full comparison of our discrete cosmological model to all the relevant astrophysical  datasets because  their independence implies that they are not all simultaneously fulfilled automatically, as for example pointed out for the inhomogeneous Lema\^itre-Tolman-Bondi model \cite{tolman}. From the mathematical perspective, one would like to understand  whether general relations among kinematical, dynamical and observational backreactions holding for any inhomogeneous solution exist because so far they must be quantified explicitly case by case.

Furthermore, it is important to quantify and clarify the role of the electric and magnetic parts of the Weyl curvature tensor on the global evolution of these discrete  universes. In our previous analysis we have shown that a nonzero second time derivative of the scale factor (this behavior would be referred to as {\it accelerating} universe in the language of the standard cosmological model)  parallel to a LRS curve is possible thanks to the electric Weyl tensor even in a spacetime without any kind of matter   \cite{cli3,cli4}. The time evolution and the magnitude of the term triggering this acceleration then depends also on the magnetic Weyl tensor \cite{cli4}. In the current analysis we have obtained a similar result showing that the electric part of the Weyl tensor affects directly the evolution of the redshift through the scale factor, while the magnetic tensor plays again only an indirect role affecting the physical quantities  only entering  the time evolution of the electric Weyl tensor.  

Thus, in this work we have shown that certain distributions of the spatial inhomogeneities within the universe which come with a nonzero Weyl curvature can nevertheless account for the supernovae dataset. In the context of inhomogeneous cosmology, to reconstruct an appropriate distribution of such spatial inhomogeneities,  which would be favored by astrophysical data, may be regarded  as  an analogous task comparable to the reconstruction of the most realistic equation of state for the cosmic fluid in the framework of the Friedman-Lemaitre-Robertson-Walker model.
 In this manner, following the line of thinking proposed in \cite{cope}, we have argued that relaxing the assumptions at the core of the Copernican principle it is possible to account for the supernovae dataset without the need of invoking any exotic fluid permeating our Universe, showing also that the distance modulus computed for some cosmological models with discretized matter content is not sensitive on how the masses are distributed within the configuration but only on the local rotational symmetry we have assumed.

\section*{Acknowledgments}
DG acknowledges support from China Postdoctoral Science Foundation (grant
No.2019M661944).

\end{document}